\documentclass[10pt,aps,floatfix,prd,onecolumn,merge,nofootinbib,
        superscriptaddress,notitlepage,longbibliography]{revtex4} 
\usepackage{graphicx} 
\usepackage{dcolumn} 
\usepackage{mathtools} 
\usepackage{slashed}
\usepackage{cancel}
\usepackage[caption=false]{subfig}
\usepackage[usenames,dvipsnames,svgnames]{xcolor}
\usepackage{braket}
\usepackage{multirow}
\usepackage{enumitem}

\def\ii{{\rm i}}

\def\e{\epsilon}

\def\ol{\overline}
\def\nl{\nonumber \\}

\def\ggto4l{g^*\!g^*\!\to\!h\!\to\!ZZ^*\!\to\!4\ell}
\def\ggTo4l{g^*\!g^*\!\to\!h\!\to\!ZZ^*\!\to\!\ell \bar\ell \ell^\prime \bar\ell^\prime}

\usepackage{hyperref} 

\usepackage[capitalise]{cleveref}

\hypersetup{%
  colorlinks=true, linktocpage=true, pdfstartpage=1, pdfstartview=FitV,%
  breaklinks=true, pdfpagemode=UseNone, pageanchor=true, pdfpagemode=UseOutlines,%
  plainpages=false, bookmarksnumbered, bookmarksopen=true, bookmarksopenlevel=1,%
  hypertexnames=true, pdfhighlight=/O,
  urlcolor=Cerulean, linkcolor=Cerulean, citecolor=Cerulean, 
}

\usepackage[normalem]{ulem}

\begin{document}
\title{$k_T$-factorization approach to the Higgs boson production in\\ $ZZ^*\to 4\ell$ channel at the LHC}
\author{Rashidul Islam}
\email{rislam@iitg.ac.in}
\affiliation{Department of Physics,
             Indian Institute of Technology Guwahati, Assam 781039, India}
\author{Mukesh Kumar}
\email{mukesh.kumar@cern.ch}
\affiliation{School of Physics and Institute for Collider Particle Physics,
             University of the Witwatersrand, Johannesburg, Wits 2050, South Africa}
\author{Vaibhav Rawoot}
\email{vaibhavrawoot@gmail.com}
\affiliation{The Institute of Mathematical Sciences,
             IV Cross Road, CIT Campus, Chennai 600 113, India}
\affiliation{{\bf Present address:} Department of Physics,
             University of Mumbai, Mumbai 400 098, India.}%
%

%
\begin{abstract}
We calculated a differential cross section of the Higgs boson production in the $h\to ZZ^*\to 4\ell$ 
decay channel within the framework of $k_T$-factorization. Results are obtained using an off-shell 
matrix element for the $g^*g^*\to h\to ZZ^*\to 4\ell$ process together with 
Ciafaloni-Catani-Fiorani-Marchesini (CCFM) evolution equations for an unintegrated gluon distribution 
function. We have presented a comparison of our results with the latest experimental measurements at 
$\sqrt{S}$ = 8~TeV and $\sqrt{S}$ = 13~TeV from the ATLAS and CMS collaborations at the LHC. In 
addition to this, we have compared our results with the results from the collinear factorization 
formalism calculated up to next-to-next-to-leading order plus next-to-next-to-leading logarithm 
(NNLO + NNLL) accuracy obtained using the HRes code for the Higgs boson production in the gluon-gluon 
fusion process. Our estimates are consistently close to NNLO + NNLL results obtained using a 
collinear factorization formalism and are also in agreement with experimental measurements.
\end{abstract}

\maketitle

%

%
\section{Introduction}
\label{sec:intro}
The Higgs boson discovery at the LHC, by ATLAS and CMS collaborations~\cite{Aad:2012tfa, Chatrchyan:2012xdj}, 
has enabled experimental measurements to be taken to investigate its properties. The ATLAS and CMS 
collaborations have performed an improved measurement of the Higgs boson mass, considering an 
invariant mass spectra of the $h\to \gamma\gamma$ and $h\to ZZ^*\to 4\ell$ decay channels~\cite{Aad:2014aba, Khachatryan:2014jba}. 
Further studies of spin and the parity quantum number of the Higgs boson have established that it is 
a neutral scalar boson, with a mass equal to 125.09 GeV, rather than a pseudoscalar 
boson~\cite{Chatrchyan:2012jja, Aad:2013xqa, Olive:2016xmw}. Its coupling strength to vector bosons 
and to fermions is studied by analyzing various decay modes of the Higgs boson~\cite{Khachatryan:2014jba, Khachatryan:2014kca, Aad:2015mxa, Aad:2015gba}. 
Establishing various aspects of the Higgs boson's properties and coupling strength allows us to study 
other aspects of it.

A dominant channel for the inclusive Higgs boson production at the LHC is gluon-gluon 
fusion~\cite{Georgi:1977gs, Graudenz:1992pv, Spira:1995rr}. Hence, the Higgs boson production at the 
LHC can be effectively used to understand the gluon dynamics inside a proton. The gluon density 
$xf_g(x,\mu^2_F)$ in a proton is a function of the Bjorken variable $x$ and the hard scale $\mu^2_F$. 
The scale evolution of parton densities, in general, is described using the 
Dokshitzer-Gribov-Lipatov-Altarelli-Parisi (DGLAP) evolution 
equation~\cite{Gribov:1972ri, Lipatov:1974qm, Altarelli:1977zs, Dokshitzer:1977sg}, where large 
logarithmic terms proportional to $\ln \mu^2_F$ are resummed up to all orders.

The factorization theorem in perturbative quantum chromodynamics (pQCD) allows us to write a 
convolution of the matrix element of the short distance process and the universal parton distribution 
functions to obtain an inclusive cross section for a given scattering process~\cite{Collins:1989gx}. 
The QCD collinear factorization theorem is based on the collinear approach where the parton 
distribution function depends on the longitudinal momentum fraction $x$ and the hard scale $\mu^2_F$. 
The Higgs boson production cross section at leading order and higher order QCD corrections to it, up 
to next-to-next-to-next-to-leading order (N$^3$LO), have been computed within the collinear 
factorization framework~\cite{Dawson:1990zj, Djouadi:1991tka, Harlander:2002wh, Anastasiou:2002yz, Ravindran:2003um, Anastasiou:2015ema}. 
However, it should be noted that the NNLO and N$^3$LO results that were obtained so far are using an 
effective theory and in the heavy top quark mass limit. Study of the Higgs boson's transverse 
momentum spectrum resummed at NNLL accuracy is shown in~\cite{deFlorian:2012mx, Bozzi:2003jy, Bozzi:2005wk}. 
Recently, the state-of-the-art predictions for the Higgs boson's transverse momentum at the LHC, at 
next-to-next-to-next-to-leading-logarithmic accuracy (N$^3$LL) matched, at NNLO is presented in 
Ref.~\cite{Bizon:2017rah}.

For the inclusive Higgs boson production at the LHC, the longitudinal momentum fraction of the 
incident gluons is small ($x_1 x_2\sim$ 0.0089-0.0175). This domain of small longitudinal momentum 
fraction ($x$) is still in the perturbative regime where is it expected that collinear factorization 
should break down because the large logarithmic term proportional to $1/x$ becomes 
dominant~\cite{Andersson:2002cf, Andersen:2003xj, Andersen:2006pg}. The contribution from the terms 
proportional to $1/x$ is taken into account in the Balitsky-Fadin-Kuraev-Lipatov (BFKL) evolution 
equation~\cite{Kuraev:1976ge, Kuraev:1977fs, Balitsky:1978ic}. An unintegrated parton densities 
(uPDFs) obeying BFKL evolution, convoluted with an off-shell matrix element within a generalized 
factorization is called $k_T$-factori\-zation~\cite{Gribov:1984tu, Levin:1991ry, Catani:1990xk, Catani:1990eg, Collins:1991ty}. 
The evolution equation is valid for both small $x$ and large $x$ is given by the 
Ciafaloni-Catani-Fiorani-Marchesini (CCFM) evolution equation~\cite{Ciafaloni:1987ur, Catani:1989yc, Catani:1989sg, Marchesini:1994wr}. 
CCFM evolution is equivalent to BFKL evolution in the limit of very small $x$ and is equivalent to 
the DGLAP evolution for a large $x$ region.

In this work, we have not implemented the reggeized parton approach~\cite{Lipatov:2000se, Bogdan:2006af, Hentschinski:2011xg, Chachamis:2012gh, Hentschinski:2011tz} 
based on the Lipatov effective action formalism, which ensures a gauge invariance of the off-shell 
amplitude~\cite{Lipatov:1995pn, Lipatov:1996ts}. However, our investigation is based on the 
assumption that the off-shell partonic amplitudes being gauge-invariant in a small-$x$ limit. 
The approach, which is based on Lipatov effective action formalism that has been employed recently in 
the calculation using $k_T$-factorization approach for the inclusive prompt photon production at 
LHC~\cite{Lipatov:2016wgr}.

In this paper, the inclusive Higgs boson production within the $k_T$-factorization approach, together 
with CCFM evolution equations have been studied and demonstrated importance of higher order 
corrections included within the $k_T$-factorization~\cite{Lipatov:2005at}. The authors of 
Ref.~\cite{Lipatov:2014mja} have shown that $k_T$-factorization gives a description of an 
experimental data from ATLAS experiment for the differential cross section of the Higgs boson 
production in the diphoton decay channel. They have calculated a leading order (LO) matrix element 
for the partonic subprocess $gg\to h\to \gamma\gamma$ considering gluons to be off-shell. Inclusive 
Higgs boson production analysis based on off-shell gluon-gluon fusion, and considering 
$H\to\gamma\gamma$, $H\to ZZ^*\to 4\ell$ (where $\ell= e,\mu$) and $H\to W^+W^- \to 
e^{\pm}\mu^{\mp}\nu\bar{\nu}$ decay channel is given in Ref.~\cite{Abdulov:2017tis}.

The ATLAS and CMS collaboration at the LHC presented a measurement of a fiducial differential cross 
section of the Higgs boson decay into four-leptons at $\sqrt{S}$ = 8~TeV~\cite{Aad:2014tca,Khachatryan:2015yvw} 
and $\sqrt{S}$ = 13~TeV~\cite{Aaboud:2017oem}. We compare the results obtained using the 
$k_T$-factorization approach with the recent ATLAS and CMS data. We have evaluated the off-shell 
matrix element for the partonic subprocess $g^*g^*\to h\to ZZ^*\to 4\ell, \ell=e,\mu$. Convolution of 
the off-shell matrix element of partonic subprocess with CCFM uPDFs~\cite{Jung:2004gs} is used to 
obtain a differential cross section.

This article is organized as follows: We discuss in detail the formalism behind our study and the 
necessary expressions for further numerical analysis in \cref{sec:formalism}. In \cref{sec:results}, 
we give the results of our numerical simulation. Here we also discuss the details of the analyses 
that have gone into the study. Finally, add we conclude and draw inferences from the 
analysis in \cref{sec:conc}.

%
\section{Formalism}
\label{sec:formalism}
In the present section, we briefly discuss the formalism we have used in our study. The 
details are in the \cref{app:amp} and \cref{app:phase_space}. In \cref{sec:intro}, we have mentioned 
that to explore the effects of $k_T$-factorization, we need to take the initial state partons 
to be off-shell. In calculating the off-shell matrix element for the process $\ggto4l$
\begin{figure}[!ht]
\centering
\resizebox{0.35\textwidth}{!}{%
  \includegraphics{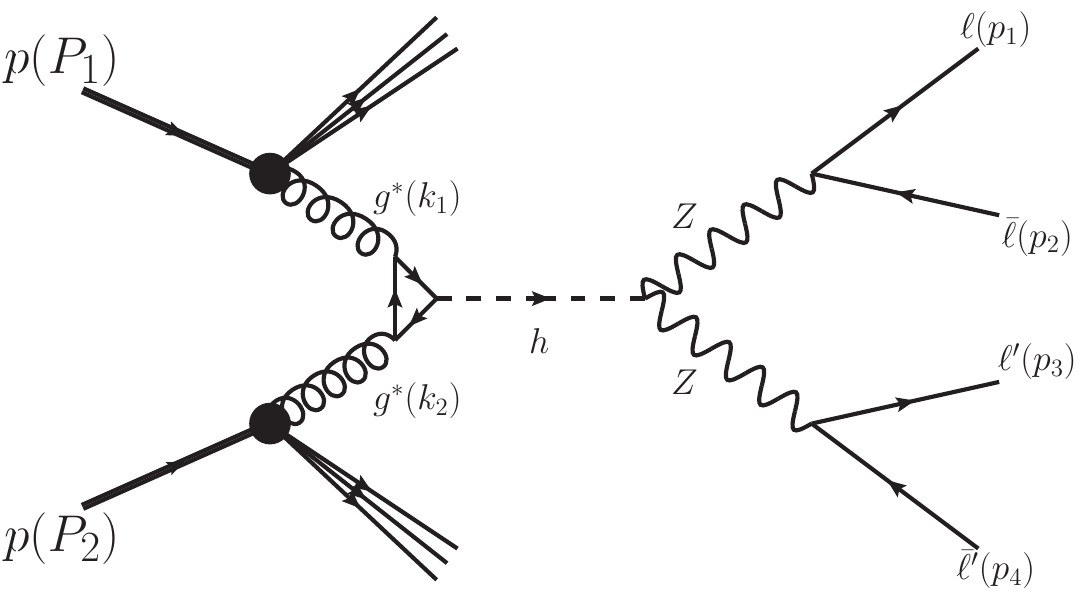}
}
\caption{Momentum assignment for the process $\ggto4l$.}
\label{fig:ZZto4l}
\end{figure}
(see Fig.~\ref{fig:ZZto4l}), we have used the effective field theory approach. The effective 
Lagrangian in the large top quark mass limit, $m_t \to \infty$, for the Higgs boson coupling to gluon 
is~\cite{Ellis:1975ap, Shifman:1979eb}
\begin{gather}
 {\cal L}_{ggh}
 =
 \frac{\alpha_s}{12 \pi} (\surd{2} G_F)^{1/2} G^a_{\mu\nu} G^{a\mu\nu} h,
 \label{lag-ggh}
\end{gather}
where $\alpha_s$ and $G_F$ are the strong and Fermi coupling constants, respectively. $G^a_{\mu\nu}$ 
is the gluon field strength tensor and $h$ is the Higgs scalar field. The effective $ggh$ triangle 
vertex (see \cref{ver-ggh2}) thus becomes
\begin{gather}
  T^{\mu\nu, ab}_{ggh} (k_1, k_2)
 =
 \delta^{ab} \frac{\alpha_S}{3 \pi} (\surd{2} G_F)^{1/2}
   [k_2^\mu k_1^\nu - (k_1 \cdot k_2) g^{\mu\nu}].
 \label{ver-ggh}
\end{gather}
The non-zero transverse momentum for an initial gluon leads to the corresponding polarization 
sum~\cite{Gribov:1984tu, Levin:1991ry}
\begin{gather}
 \ol{\sum} \e^{*k}_\mu \e^{k}_\nu
 \simeq
 \frac{k_{\perp\mu} k_{\perp\nu}}{k^2_{\perp}}.
 \label{pol-sum}
\end{gather}
Using \cref{ver-ggh,pol-sum}, we derived the off-shell matrix element for the hard scattering 
process. The matrix element thus obtained is given by \cref{amp_higgs6} as
\begin{gather}
 \ol{|{\cal M}|^2}
 =
 \frac{2}{9} \frac{\alpha^2_S}{\pi^2} \frac{m^4_Z}{v^4}
 \frac{[\hat s + (\sum_{i=1}^2 {\bf k}_{i\perp})^2]^2}
      {[(\hat s - m^2_h)^2 - m^2_h \Gamma^2_h ]}
 \cos^2\varphi
 \ \times \frac{[g^4_L + g^4_R] (p_1 \cdot p_3) (p_2 \cdot p_4)
       + 2 g^2_L g^2_R (p_1 \cdot p_4) (p_2 \cdot p_3)}
        {[(2 p_1 \cdot p_2 - m^2_Z)^2 + m^2_Z \Gamma^2_Z]
       \ [(2 p_3 \cdot p_4 - m^2_Z)^2 + m^2_Z \Gamma^2_Z]},
  \label{offshellamp}
\end{gather}
where
\begin{align*}
 g_L =& \frac{g_W}{\cos\theta_W} \Big( - \frac{1}{2} + \sin^2\theta_W \Big),
 \\
 g_R =& \frac{g_W}{\cos\theta_W} \sin^2\theta_W, & \text{and} \quad
 v=(\surd{2}G_F)^{-1/2}.
\end{align*}
Here $\Gamma_h$ and $\Gamma_Z$ are the total decay widths of the Higgs boson and $Z$ boson, 
respectively. ${\bf k}_{i\perp}\!\!$ are the intrinsic transverse momenta of the initial gluons. 
$\varphi$ is the azimuthal angle between ${\bf k}_{1\perp}$ and ${\bf k}_{2\perp}$. $m_h$ and $m_Z$ 
are the Higgs boson and $Z$ boson masses, respectively. The partonic centre of mass energy 
is denoted by $\hat s$. $\theta_W$ and $g_W$ are the weak mixing angle and the coupling of weak 
interaction, respectively.

Finally, we arrived at the hadronic cross section for the off-shell hard scattering amplitude of 
\cref{offshellamp} within the framework of $k_T$-factorization as~\footnote[1]{The processes 
contributing to $4$ lepton final states are $e^+e^-e^+e^-$, $\mu^+\mu^-\mu^+\mu^-$, $e^+e^-\mu^+\mu^-$ 
and $\mu^+\mu^-e^+e^-$. We have added these contributions with proper weight. The first two 
processes have two pairs of identical particles. Hence their phase space has to be multiplied by a 
factor $(1/2)\times(1/2) = 1/4$ to get rid of overcounting.} (see \cref{factorization4})
\begin{gather}
 \sigma
 =
 \!\!\int\!\!\prod_{i=1}^2 \frac{f_g(x_i,{\bf k}^2_{i\perp},\mu^2_F)}{x^2_i S^2}
 d{\bf k}^2_{i\perp} \frac{d\varphi_i}{2 \pi}
 \prod_{f=1}^3 d^2{\bf p}_{f\perp} dy_f dy_4
 \frac{\ol{|{\cal M}|^2}}{2^{12}\ \pi^5},
 \label{facttmd5}
\end{gather}
with the longitudinal momentum fractions $x_1$ and $x_2$ of initial gluons to be
\begin{gather}
  x_1 = \sum_{f=1}^4\frac{|{\bf p}_{f\perp}|}{\sqrt{S}} e^{y_f},
  \qquad
  x_2 = \sum_{f=1}^4\frac{|{\bf p}_{f\perp}|}{\sqrt{S}} e^{-y_f},
\end{gather}
and the transverse momenta
\begin{gather}
 \sum_{i=1}^2 {\bf k}_{i\perp} = \sum_{f=1}^4 {\bf p}_{f\perp}.
\end{gather}
In \cref{facttmd5}, $\varphi_{1,2}$ are the azimuthal angle of ${\bf k}_{1\perp,2\perp}$. $y$ and 
${\bf p}_{\perp}\!\!$ are rapidities and transverse momenta of the final state leptons, respectively. 
The hadronic centre of mass energy is denoted by $S$.

%
\section{Results and Discussion}
\label{sec:results}
With all the calculational tools at our disposal, we proceed to perform a numerical calculation using 
\cref{facttmd5} together with the off-shell hard scattering amplitude given in \cref{offshellamp}. We 
estimate the cross section of the Higgs boson production as a function of transverse momentum ($p_T$) 
and rapidity ($y$) of the Higgs boson in the four-lepton decay channel. Results are obtained using 
CCFM A0 the set of uPDFs~\cite{Jung:2004gs} which is commonly used for such phenomenological studies. 
Recently a fit to a high precision data from deeply inelastic scattering at the HERA is performed 
using a $k_T$-factorization and CCFM evolution~\cite{Hautmann:2013tba}. A transverse momentum 
dependent gluon density function including experimental and theoretical uncertainties were 
obtained. The application of these unintegrated gluon densities to vector boson + jet production 
processes at LHC is given in Ref.~\cite{Dooling:2014kia}. Unintegrated gluon densities including 
experimental and theoretical uncertainties are given in the CCFM JH2013-set in TMDlib 
library~\cite{Hautmann:2014kza,Connor:2016bmt}. We have used the TMDlib library to calculate our 
results using CCFM JH2013-set. For our phenomenological study, we have used the CCFM JH2013-set2 
which is determined from the fit to both structure functions $F^{(charm)}_2$ and $F_2$ data 
whereas CCFM JH2013-set1 is determined from the fit to inclusive $F_2$ data only.

Total decay width and mass of the Higgs boson is set to be equal to 4.0~MeV and 125.09~GeV, 
respectively~\cite{Olive:2016xmw}. We have implemented kinematical cuts on the rapidity 
and transverse momentum of leptons used by ATLAS and CMS experiments in their measurements. For the 
ATLAS experiment, the absolute value of rapidity is $|\eta|< 2.5$ and the leading transverse momentum 
of the lepton is $p_T < 20$~GeV. The transverse momenta of sub-leading leptons are $p_T < 15, 10, 
7$~GeV. Similarly for the CMS experiment, the absolute value of rapidity is $|\eta|< 2.5$ and the 
ordered transverse momenta of the leptons are $p_T < 20, 10, 7, 7$~GeV. The cross section in 
\cref{facttmd5} depends on the renormalization and factorization scales $\mu_R$ and $\mu_F$, 
respectively. The scale uncertainty in the cross section is estimated by varying the scale between 
$\mu_R = \mu_F = m_h/2$ and $\mu_R = \mu_F = 2 m_h$.

We have also calculated a total inclusive cross section for the Higgs boson production with $k_T \to 0$ 
and averaged over an azimuthal angle of the Higgs boson. This result of the inclusive cross 
section is equivalent to the cross section obtained using the collinear factorization approach at LO 
while using collinear parton densities. We have used the Martin-Stirling-Thorne-Watt (MSTW) 
set~\cite{Martin:2009iq} for collinear parton densities. We have also obtained results of the total 
inclusive cross section for the Higgs production with $k_T$-factorization formalism using the CCFM 
JH2013-set2 of uPDFs.
\begin{table}[!ht]
\centering
\begin{tabular}{ccc}
\hline
\multirow{2}*{$\sqrt{S}$ (TeV)} & \multicolumn{2}{c}{$\sigma^{tot}$ (pb)}  \\
                                & ($k_T$-factorization)& (Collinear factorization) \\
\hline \hline
8 & 27.12 & 6.11   \\
13 & 66.40 & 15.10  \\
\hline
\end{tabular}
\caption{Total inclusive cross section ($\sigma^{tot}$) for the Higgs boson
         production in gluon-gluon fusion channel.}
\label{table1}
\end{table}

In \cref{table1}, we have given our results for the total inclusive cross section for the Higgs 
production, using both collinear and $k_T$-factorization framework. Our results, for total inclusive 
cross section, obtained using collinear approach, are consistent with the results obtained in 
Ref.~\cite{Anastasiou:2012hx} at $\sqrt{S}$ = 8~TeV. The results obtained with $k_T$-factorization is 
close to next-to-leading order (NNLO) results given in Ref.~\cite{Anastasiou:2012hx} at $\sqrt{S}$ = 
8~TeV. The cross section estimates given here are for a gluon-gluon fusion process only. The 
inclusive cross section for the Higgs boson production can be obtained using a hadron level Monte 
Carlo event generator called CASCADE~\cite{Jung:2010si}. CASCADE uses the CCFM evolution equation in 
the initial state with the off-shell parton level matrix element.

\begin{figure}[!ht]
 \centering
 \includegraphics[trim=0 0 0 0,clip,width=0.45\linewidth,height=0.4\linewidth]
            {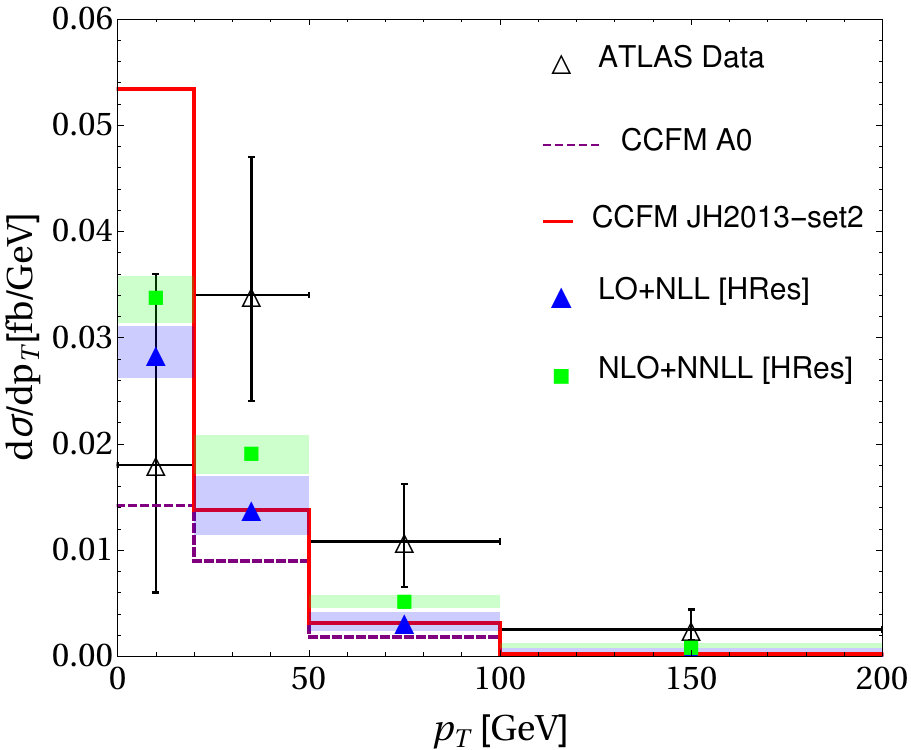}
 \includegraphics[trim=0 0 0 0,clip,width=0.45\linewidth,height=0.4\linewidth]
            {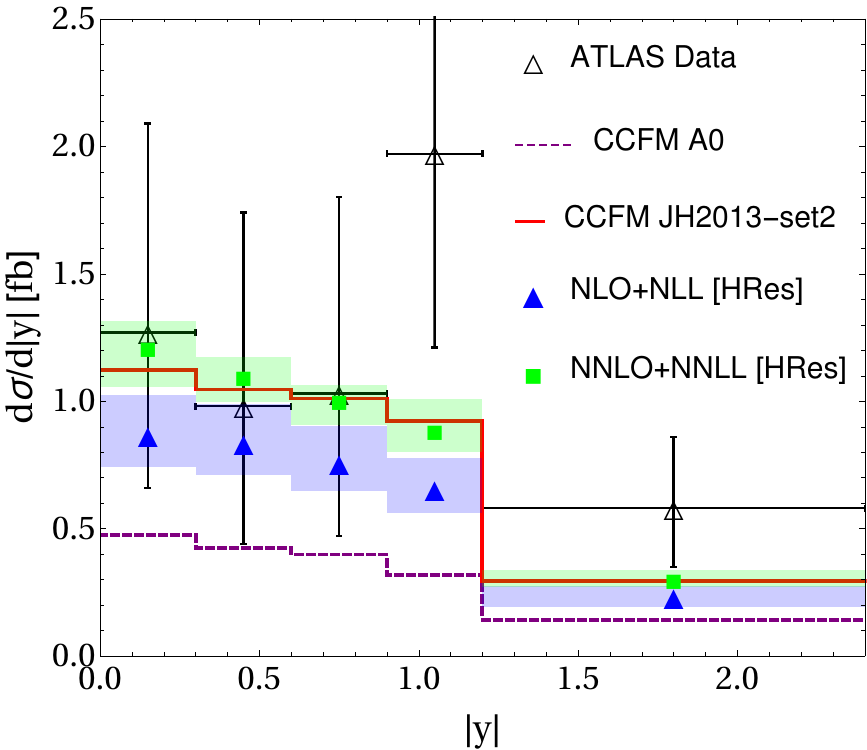}
 \caption{Differential cross section of the Higgs boson production as a function of transverse 
          momentum ($p_T$) and rapidity ($y$) of the Higgs boson in four-lepton decay channel at 
          $\sqrt{S}$ = 8~TeV. Solid (red) line and dashed (purple) line is a result obtained 
          using $k_T$-factorization approach with CCFM JH2013-set2 and CCFM A0 unintegrated gluon 
          densities respectively. Filled triangle and filled square points corresponds to estimates 
          obtained using HRes tool~\cite{deFlorian:2012mx, Grazzini:2013mca} up to NNLO + NNLL 
          accuracy and shaded region corresponds to scale uncertainty in renormalization and 
          factorization scale. Experimental data points are from ATLAS~\cite{Aad:2014tca}. The error 
          bars on the data points shows total (statistical $\oplus$ systematic) uncertainty.}
\label{fig1}
\end{figure}
\begin{figure}[!ht]
 \centering
 \includegraphics[trim=0 0 0 0,clip,width=0.45\linewidth,height=0.4\linewidth]
            {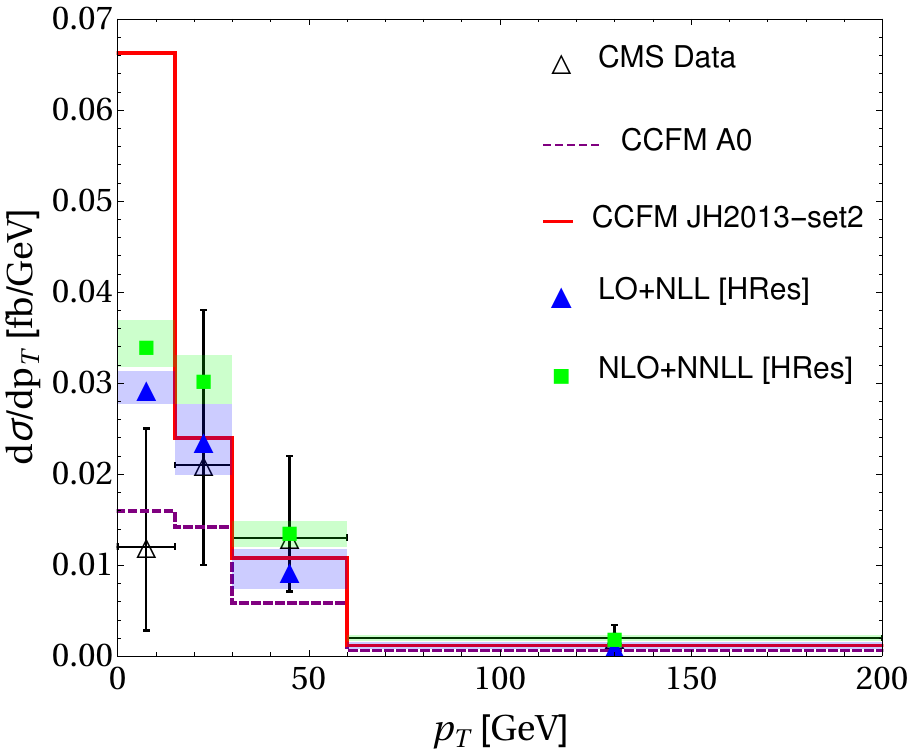}
 \includegraphics[trim=0 0 0 0,clip,width=0.45\linewidth,height=0.4\linewidth]
            {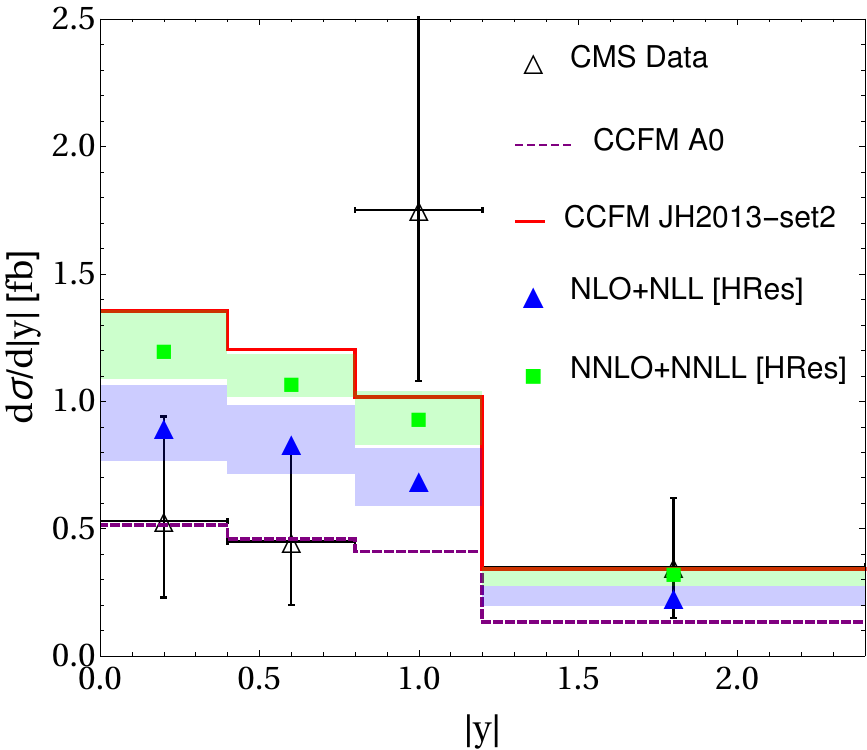}
 \caption{Differential cross section of the Higgs boson production as a function of transverse 
          momentum ($p_T$) and rapidity ($y$) of the Higgs boson in four-lepton decay channel at 
          $\sqrt{S}$ = 8~TeV. Notations of all the histograms are the same as in 
          \cref{fig1}. Higher order pQCD predictions up to NNLO + NNLL accuracy are obtained using 
          HRes tool~\cite{deFlorian:2012mx,Grazzini:2013mca}. Experimental data points are from 
          CMS~\cite{Khachatryan:2015yvw}.}
\label{fig2}
\end{figure}
\begin{figure}[!ht]
 \centering
 \includegraphics[trim=0 0 0 0,clip,width=0.45\linewidth,height=0.4\linewidth]
            {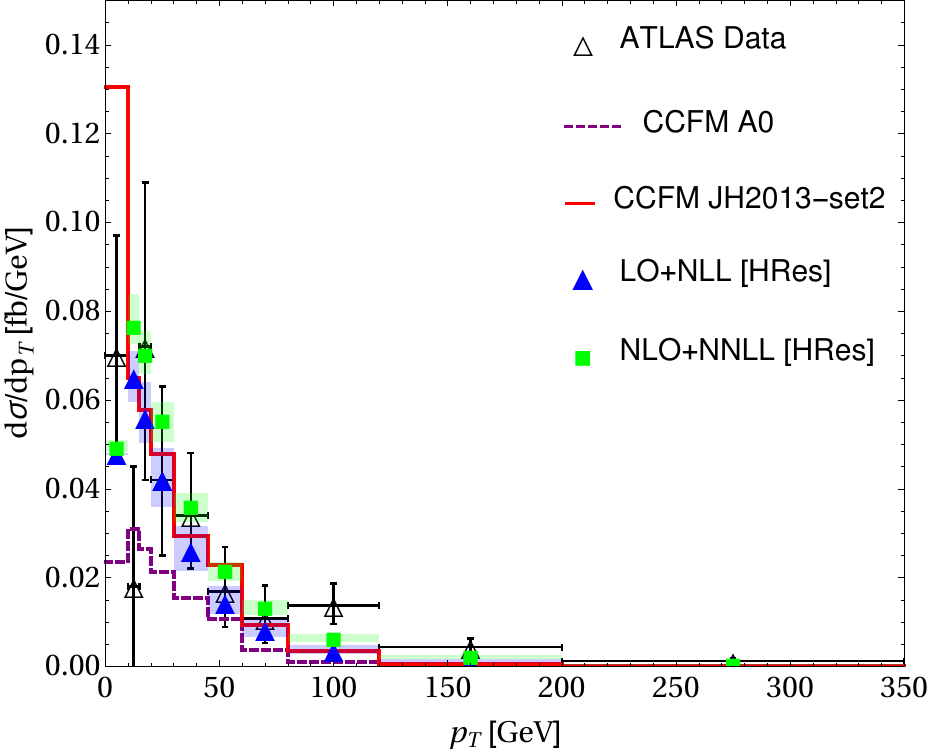}
 \includegraphics[trim=0 0 0 0,clip,width=0.45\linewidth,height=0.4\linewidth]
            {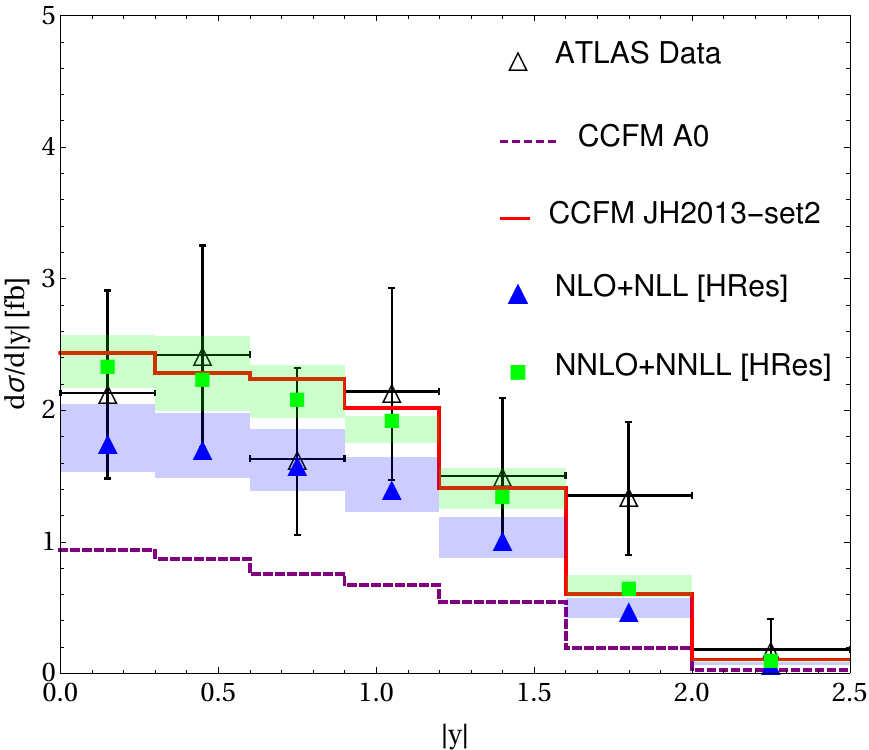}
 \caption{Differential cross section of the Higgs boson production as a function of transverse 
          momentum ($p_T$) and rapidity ($y$) of the Higgs boson in four-lepton decay channel at 
          $\sqrt{S}$ = 13~TeV. Notations of all the histograms are the same as in 
          \cref{fig1}. Higher order pQCD predictions up to NNLO + NNLL accuracy are obtained using 
          HRes tool~\cite{deFlorian:2012mx, Grazzini:2013mca}.}
\label{fig3}
\end{figure}

We have presented our results in \cref{fig1,fig2,fig3}. \cref{fig1,fig2,fig3} shows the result of the 
differential cross section for the Higgs boson production in the four-lepton decay channel at $
\sqrt{S}$ = 8~TeV and $\sqrt{S}$ = 13~TeV, respectively. We have compared our results of the 
differential cross section obtained using the $k_T$-factorization approach with experimental 
measurements from the ATLAS at $\sqrt{S}$ = 8~TeV, 13 TeV and CMS collaboration at $\sqrt{S}$ = 
8~TeV~\cite{Aad:2014tca, Khachatryan:2015yvw}. The solid (red) and dashed (purple) histogram 
corresponds to our results obtained using the CCFM JH2013-set2 set of uPDFs and CCFM A0 set of uPDFs 
respectively. We also see that the results obtained using CCFM JH2013-set2 has a better agreement 
with experimetal measurements than A0 set.

Our results are plotted against state-of-the-art results for the cross section calculated up to 
next-to-leading order plus next-to-leading logarithm (NLO + NLL) and next-to-next-to-leading order 
plus next-to-next-leading logarithm (NNLO + NNLL) obtained using the HRes tool~\cite{deFlorian:2012mx, Grazzini:2013mca} 
within the collinear factorization framework. Our results of both the differential cross section in 
$p_T$ and $y$ using the $k_T$-factoriza\-tion framework with CCFM unintegrated PDFs are consistently 
close to NNLO + NNLL results at $\sqrt{S}$ = 8~TeV and $\sqrt{S}$ = 13~TeV. This can be explained 
considering the fact that the main part of higher order corrections are included in the $k_T$-
factorization approach~\cite{Lipatov:2005at,Hautmann:2002tu,Ryskin:1999yq}. For the $p_T$ 
distribution, we are using the convention that NLO + NLL and NNLO + NNLL results are labelled as 
LO + NLL and NLO + NNLL, respectively considering that the $p_T$ distribution is non-zero at NLO.

%
\section{Conclusions}
\label{sec:conc}
In this paper, add we present a phenomenological study of the Higgs boson production in the 
four-lepton decay channel within the $k_T$-factorization framework. Here CCFM unintegrated 
parton densities were convoluted with the hard matrix element, considering initial gluons to be off-
shell. We present a comparison of our results with experimental measurements. Our results are 
evaluated using same experimental conditions (i.e., the same $p_T$ and $y$ cuts were used for our 
estimates as given by the experimental results) for both the ATLAS and CMS at $\sqrt{S}$ = 8~TeV and 
at $\sqrt{S}$ = 13~TeV, respectively.

Further comparison of our estimates with the state-of-the-art results of a differential cross section 
within collinear factorization up to NLO + NLL and NNLO + NNLL obtained using HRes code is presented. 
We have also estimated a total inclusive cross section for the Higgs boson production within both 
collinear factorization and $k_T$-factorization framework. Within $k_T$-factorization approach, 
we have compared the results obtained using CCFM JH2013-set2 and CCFM A0 uPDF set. Our results for 
the differential cross section with CCFM JH2012-set2 are close to the NNLO + NNLL results obtained 
using the HRes tool.

Our results show that the observed $p_T$ distribution of the final state can be generated at leading 
order subprocesses, using unintegrated gluon distributions. Moreover, gluons in the initial state 
have finite transverse momenta, which results in the transverse momenta of the final state. 
The total inclusive cross section estimated using $k_T$-factorization is close to the NNLO results 
obtained using collinear factorization. The main reason for this behavior is that the main part of 
higher order correction in collinear pQCD is already included in the $k_T$-factorization~\cite{Watt:2003vf} 
framework.

The higher order corrections within $k_T$-factorization at the parton level would be an interesting 
study, to see any additional effect. The cross section for the Higgs boson production has been 
calculated using a mixture of LO and NLO partonic diagrams and unintegrated PDFs from the $k_T$-
factorization. Considering the effect of the transverse momentum of the initial gluon on the 
transverse momentum distribution of the final state, our study, add as well as further studies 
in this direction could impose constraints on uPDFs of gluons.

%
\appendix

\section{Amplitude of \texorpdfstring{$\ggto4l$}{g*g* -> h -> ZZ* -> 4l}}
\label{app:amp}
In this appendix, add we give details of our calculations that went into our analysis. \cref{fig:ZZto4l} 
shows the assignment of the momenta. In the limit of large top quark mass $m_t \to \infty$, the 
effective Lagrangian for the Higgs boson coupling to gluons given in \cref{lag-ggh} can be written as
\begin{align}
 {\cal L}_{ggh}
 =
 \frac{\alpha_s}{12 \pi v} G^a_{\mu\nu} G^{a\mu\nu} h,
 &\quad
 [ \text{Using } v = (\surd{2} G_F)^{-1/2} ]
 \label{lag_gluon}
\end{align}
where $v$ is the vacuum expectation value of the scalar field. The amplitude of the process $\ggTo4l$ is
\begin{align}
 \ii {\cal M}
 =
 \e^{*k_1}_{\mu} \e^{*k_2}_{\nu}
 \frac{[\ii T^{\mu\nu, ab}_{ggh} (k_1, k_2)] \ii
 [\ii T_{h \to ZZ^* \to 4\ell} (p_1, p_2, p_3, p_4)]}
      {(\hat s - m^2_h) + i m_h \Gamma_h},
 \label{amp_higgs3}
\end{align}
where
\begin{gather}
 T^{\mu\nu, ab}_{ggh} (k_1, k_2)
 =
 \delta^{ab} \frac{\alpha_s}{3 \pi v} [k_2^\mu k_1^\nu - (k_1 \cdot k_2) g^{\mu\nu}],
 \label{ver-ggh2}
 \\
 T_{h \to ZZ^* \to 4\ell} (p_1, p_2, p_3, p_4)
 =
 - 2 \frac{m^2_Z}{v}
 \ \frac{\bar u(p_1) \gamma^\rho [g^{(1)}_L P_L + g^{(1)}_R P_R] v(p_2)}
        {[(p_1 + p_2)^2 - m^2_Z + i m_Z \Gamma_Z]}
  \times
   \frac{\bar u(p_3) \gamma_\rho [g^{(2)}_L P_L + g^{(2)}_R P_R] v(p_4)}
        {[(p_3 + p_4)^2 - m^2_Z + i m_Z \Gamma_Z]},
\end{gather}
where
\begin{gather}
 g^{(i)}_L = \frac{g_W}{c_W} \Big( - \frac{1}{2} + s^2_W \Big),
 \quad
 g^{(i)}_R = \frac{g_W}{c_W} s^2_W,
 \quad
 i = 1,2.
\end{gather}
Hence we shall denote them only by $g_L$ and $g_R$ hereafter. 

Now using the above expressions into \cref{amp_higgs3} we can get $\ol{|{\cal M}|^2}$ for the process as
\begin{align}
 \ol{|{\cal M}|^2}
 =&
 |T^{\mu\nu, ab}_{ggh} (k_1, k_2)|^2
 \frac{|T_{h \to ZZ^* \to 4\ell} (p_1, p_2, p_3, p_4)|^2}
      {(\hat s - m^2_h)^2 + m^2_h \Gamma^2_h},
\end{align}
where
\begin{gather}
 |T_{h \to ZZ^* \to 4\ell} (p_1, p_2, p_3, p_4)|^2
 =
 64 \frac{m^4_Z}{v^2}
 \ \times \frac{[g^4_L + g^4_R] (p_1 \cdot p_3) (p_2 \cdot p_4)
       + 2 g^2_L g^2_R (p_1 \cdot p_4) (p_2 \cdot p_3)}
        {[(2 p_1 \cdot p_2 - m^2_Z)^2 + m^2_Z \Gamma^2_Z]
       \ [(2 p_3 \cdot p_4 - m^2_Z)^2 + m^2_Z \Gamma^2_Z]},
\end{gather}
whereas
\begin{gather}
 |T^{\mu\nu, ab}_{ggh} (k_1, k_2)|^2
 =
 \ol{\delta^{ab}\delta^{ab}} \frac{1}{9} \frac{\alpha^2_s}{\pi^2} \frac{1}{v^2}
 \times
 \ol{\sum} \e^{*k_1}_{\mu}\e^{k_1}_{\mu^\prime} \ \ol{\sum} \e^{*k_2}_{\nu}\e^{k_2}_{\nu^\prime}
 [k_2^\mu k_1^\nu - (k_1 \cdot k_2) g^{\mu\nu}]
 \times
     \ [k_2^{\mu^\prime} k_1^{\nu^\prime} - (k_1 \cdot k_2) g^{\mu^\prime\nu^\prime}].
 \label{amp_higgs5}
\end{gather}
To calculate the gluon part of the above amplitude we use the Sudakov decomposition of momenta as 
followed in Ref.~\cite{Catani:1990eg, Collins:1991ty}. Accordingly, we can write the gluon momenta 
$k_{1,2}$ as
\begin{align}
 \begin{split}
  k_1 =& z_1 P_1 + \bar z_1 P_2 + k_{1\perp},
  \\
  k_2 =& z_2 P_1 + \bar z_2 P_2 + k_{2\perp},
 \end{split}
 \quad \text{where}
 \begin{split}\label{def_kT}
  k_{1\perp} = (0, {\bf k}_{1\perp}, 0),
  \\
  k_{2\perp} = (0, {\bf k}_{2\perp}, 0).
 \end{split}
\end{align}
In the above equation, $k_{1\perp,2\perp}$ are vectors transverse to the momenta $P_{1,2}$ of the 
incoming hadrons. It is convenient to take both $P_{1,2}$ to be light-like (i.e. $P^2_{1,2} = 0$). 
Because of this, $P_{1,2}$ are slightly different from the actual momenta of the incoming hadrons. 
Let us take $P_{1,2}$ to be as follows
\begin{align}
 P_{1,2} = \frac{\sqrt{S}}{2} (1, {\bf 0}, \pm 1),
 \label{mom_had}
\end{align}
where $\sqrt{S}$ is the centre of mass energy of the hadrons. In the high energy limit, the introduction of strong ordering in longitudinal momenta gives
\begin{align}
 \begin{split}\label{mom_part}
  k_1 =& z_1 P_1 + k_{1\perp},
  \\
  k_2 =& \bar z_2 P_2 + k_{2\perp}.
 \end{split}
\end{align}
Therefore using \cref{mom_had,mom_part} we get
\begin{gather}
 k_1 \cdot k_2
 =
 z_1 \bar z_2 P_1 \cdot P_2 + k_{1\perp} \cdot k_{2\perp},
 \label{def_k1dotk2}
 \\
 \hat s
 =
 (k_1 + k_2)^2
 =
 2 z_1 \bar z_2 P_1 \cdot P_2 - ({\bf k}_{1\perp} + {\bf k}_{2\perp})^2.
 \label{def_shat}
\end{gather}
In the last step, add we have used the definition of $k_{1\perp,2\perp}$ introduced in \cref{def_kT}.

With the help of the above expressions, we are now ready to calculate the gluon part of \cref{amp_higgs5}. Using the polarization sum for the off-shell gluons given by
\begin{gather}
 \ol{\sum} \e^{*k}_\mu \e^{k}_\nu
 \simeq
 \frac{k_{\perp\mu} k_{\perp\nu}}{k^2_{\perp}},
\end{gather}
we can calculate
\begin{align}
 \ol{\sum} \e^{*k_1}_{\mu}\e^{k_1}_{\mu^\prime}
 &
 \ol{\sum} \e^{*k_2}_{\nu}\e^{k_2}_{\nu^\prime}
 [k_2^\mu k_1^\nu - (k_1 \cdot k_2) g^{\mu\nu}]
 \times
 \ [k_2^{\mu^\prime} k_1^{\nu^\prime} - (k_1 \cdot k_2) g^{\mu^\prime\nu^\prime}]
 \nl
 =&
 \frac{1}{4}
 [\hat s + ({\bf k}_{1\perp} + {\bf k}_{2\perp})^2]^2
 \cos^2\varphi.
 &
 [\text{Using Eq.~\eqref{def_k1dotk2} and then Eq.~\eqref{def_shat}}]
 \label{amp_gluon_off-shell}
\end{align}
Therefore putting the expression of \cref{amp_gluon_off-shell} into \cref{amp_higgs5} we get the 
total spin averaged squared amplitude of the process $\ggTo4l$ as
\begin{gather}
 \ol{|{\cal M}|^2}
 =
 \frac{2}{9} \frac{\alpha^2_S}{\pi^2} \frac{m^4_Z}{v^4}
 \frac{[\hat s + ({\bf k}_{1\perp} + {\bf k}_{2\perp})^2]^2}
      {(\hat s - m^2_h)^2 + m^2_h \Gamma^2_h}
 \cos^2\varphi
 \ \times \frac{[g^4_L + g^4_R] (p_1 \cdot p_3) (p_2 \cdot p_4)
       + 2 g^2_L g^2_R (p_1 \cdot p_4) (p_2 \cdot p_3)}
        {[(2 p_1 \cdot p_2 - m^2_Z)^2 + m^2_Z \Gamma^2_Z]
       \ [(2 p_3 \cdot p_4 - m^2_Z)^2 + m^2_Z \Gamma^2_Z]}.
 \label{amp_higgs6}
\end{gather}


\section{Phase Space Calculation}
\label{app:phase_space}
In this appendix, we give details of the phase space calculations related to our analysis. 
Let us express the 4-momentum $p$ in a 3-component vector in terms of the transverse momentum 
${\bf p}_T$ and the rapidity $y$ as follows
\begin{gather}
 p = (E_T \cosh y, {\bf p}_T, E_T \sinh y),
\end{gather}
where the transverse energy $E_T = \sqrt{{\bf p}^2_T + m^2}$ and $m$ is the mass. We can write the 
measure of the phase space integration as
\begin{gather}
 \frac{d^3{\bf p}}{p_{_0}}
 =
 d^2{\bf p}_T dy.
\end{gather}
Let us further express the 4-momentum $p$ in a 4-component vector in terms of $p_T (\equiv |{\bf p}_T|)$, 
$y$ and the azimuthal angle $\varphi$ as
\begin{gather}
 p = (E_T \cosh y, p_T \cos\varphi, p_T \sin\varphi, E_T \sinh y).
\end{gather}
Now the measure of integration over the phase space becomes
\begin{gather}
 \frac{d^3{\bf p}}{p_{_0}}
 =
 p_T dp_T d\varphi dy.
\end{gather}
The above can be easily understood if we express the 2-component vector ${\bf p}_T$ in polar 
coordinates $(p_T \cos\varphi, p_T \sin\varphi)$. Then we can write $d^2{\bf p}_T = p_T dp_T d\varphi$.

Now let us turn our attention to the phase space integration of a 4-body final state process in a 
hadron collider. The hadronic cross section in the $k_T$-factorization approach for the off-shell 
hard scattering amplitude of \cref{app:amp} is (see Eq.~2.1 of Ref.~\cite{Szczurek:2014mwa})
\begin{gather}
 \sigma
 =
 \!\!\int\!\!\frac{1}{x_1 x_2 S} \prod_{i=1}^2 \frac{dx_i}{x_i}
 f_g(x_i,{\bf k}^2_{i\perp},\mu^2_F) \frac{d^2{\bf k}_{i\perp}}{\pi}
 \times
 \prod_{f=1}^4 \frac{d^3{\bf p}_f}{(2\pi)^3 2p_{f0}}
 \ol{|{\cal M}|^2}
 \times
 (2\pi)^4 \delta^{(4)} \bigg(\sum_{i=1}^2 k_i - \sum_{f=1}^4 p_i\bigg).
 \label{factorization}
\end{gather}
Introducing the azimuthal angles $\varphi_{1,2}$ of the off-shell gluons into \cref{factorization} we 
can write the gluon phase space as
\begin{gather}
 \int\!\!\frac{d^2{\bf k}_{i\perp}}{\pi}
 =
 \int\!\!\frac{k_{i\perp} dk_{i\perp} d\varphi_i}{\pi}
 =
 \int\!\!d{\bf k}^2_{i\perp} \frac{d\varphi_i}{2 \pi}.
 \label{gluon_ps}
\end{gather}
Let us write the 4-momenta of the partons, using \cref{mom_part}, as
\begin{align}
\begin{split}
 k_1 =& \bigg( x_1 \frac{\sqrt{S}}{2}, {\bf k}_{1\perp},   x_1 \frac{\sqrt{S}}{2} \bigg),
 \\
 k_2 =& \bigg( x_2 \frac{\sqrt{S}}{2}, {\bf k}_{2\perp}, - x_2 \frac{\sqrt{S}}{2} \bigg).
\end{split}
\end{align}
Here $z_1 \to x_1$ and $\bar z_2 \to x_2$. Also the 4-momenta of the leptons are
\begin{gather}
 \begin{gathered}
  p_f
  = (|{\bf p}_{f\perp}| \cosh y_f, {\bf p}_{f\perp}, |{\bf p}_{f\perp}| \sinh y_f).
 \end{gathered}
\end{gather}
Using the above designation of the gluon and lepton momenta we can simplify the delta function of 
\cref{factorization} as
\begin{gather}
 \delta^{(4)} \bigg(\sum_{i=1}^2 k_i - \sum_{f=1}^4 p_i\bigg)
 =
 \frac{1}{S}
 \delta^{(2)} \bigg( \sum_{i=1}^2 {\bf k}_{i\perp} - \sum_{f=1}^4 {\bf p}_{f\perp} \bigg)
 \times
 \delta \bigg( x_1 - \sum_{f=1}^4\frac{|{\bf p}_{f\perp}|}{\sqrt{S}} e^{y_f} \bigg)
 \ \delta \bigg( x_2 - \sum_{f=1}^4\frac{|{\bf p}_{f\perp}|}{\sqrt{S}} e^{-y_f} \bigg).
 \label{delta_fn}
\end{gather}
Putting the results from \cref{gluon_ps,delta_fn} into \cref{factorization} and integrating over 
$x_1$ and $x_2$ we get
\begin{gather}
 \sigma
 =
 \!\!\int\!\!\prod_{i=1}^2 \frac{f_g(x_i,{\bf k}^2_{i\perp},\mu^2_F)}{x^2_i S^2}
 d{\bf k}^2_{i\perp} \frac{d\varphi_i}{2 \pi}
 \times
 \prod_{f=1}^4 d^2{\bf p}_{f\perp} dy_f
 \frac{\ol{|{\cal M}|^2}}{2^{12}\ \pi^8}
 \ \delta^{(2)} \bigg( \sum_{i=1}^2 {\bf k}_{i\perp} - \sum_{f=1}^4 {\bf p}_{f\perp} \bigg).
 \label{factorization3}
\end{gather}
where
\begin{gather}
  x_1 = \sum_{f=1}^4\frac{|{\bf p}_{f\perp}|}{\sqrt{S}} e^{y_f},
  \qquad
  x_2 = \sum_{f=1}^4\frac{|{\bf p}_{f\perp}|}{\sqrt{S}} e^{-y_f},
 \\
 \sum_{i=1}^2 {\bf k}_{i\perp} = \sum_{f=1}^4 {\bf p}_{f\perp}.
\end{gather}
Now integration over ${\bf p}_{4\perp}$ in \cref{factorization3} gives us
\begin{gather}
 \sigma
 =
 \!\!\int\!\!\prod_{i=1}^2 \frac{f_g(x_i,{\bf k}^2_{i\perp},\mu^2_F)}{x^2_i S^2}
 d{\bf k}^2_{i\perp} \frac{d\varphi_i}{2 \pi}
 \prod_{f=1}^3 d^2{\bf p}_{f\perp} dy_f dy_4
 \frac{\ol{|{\cal M}|^2}}{2^{12}\ \pi^5}.
 \label{factorization4}
\end{gather}
In the last step above we have used $d^2{\bf p}_{f\perp} = 2\pi p_{f\perp} dp_{f\perp} = \pi d{\bf p}^2_{f\perp}, (f=1,2,3)$.

%

\section*{Acknowledgements}

The authors would like to extend their sincere gratitude to V. Ravindran for continuous support and 
fruitful discussions during the course of this work and A. Lipatov for providing their codes of a 
similar study for diphoton distributions. VR would like to thank Giancarlo Ferrera for useful 
discussions on HRes tool and related topics. RI would also like to acknowledge the hospitality 
provided by the Institute of Mathematical Sciences, Chennai, India where a part of the work has been 
done. RI thank the SERB-DST, India for the research grant EMR/2015/000333.

\bibliographystyle{apsrev4-1}
\bibliography{TMD.bib}
\end{document}